\documentclass[twocolumn,showkeys]{revtex4}
\usepackage[dvips]{epsfig}
\usepackage{natbib,amsmath,amsfonts}

\newcommand{\eg}{{\em e.g.}\/ }
\newcommand{\ie}{{\em i.e.}\/ }
\newcommand{\xmin}{x_\mathrm{min}}

\begin{document}

\title{On fitting power laws to ecological data}

\author{A. James}
\email{a.james@math.canterbury.ac.nz}
 \author{M. J. Plank}
 \email{m.plank@math.canterbury.ac.nz}
\affiliation{Biomathematics Research Centre, University of Canterbury, Christchurch, New Zealand.}

\begin{abstract}
Heavy-tailed or power-law distributions are becoming increasingly common in biological literature. A wide range of biological data has been fitted to distributions with heavy tails. Many of these studies use simple fitting methods to find the parameters in the distribution, which can give highly misleading results. The potential pitfalls that can occur when using these methods are pointed out, and a step-by-step guide to fitting power-law distributions and assessing their goodness-of-fit is offered.
\end{abstract}

\maketitle

\section{Introduction}
In the last fifteen years, power-law distributions, with their heavy-tailed and scale-invariant features, have become ubiquitous in scientific literature in general \citep{newman:06,sims:07}.
Heavy-tailed distributions have been fitted to data from a wide range of sources, including ecological size spectra \citep{vidondo:97}, dispersal functions for spores \citep{sundberg:05}, seeds \citep{katul:05} and birds \citep{paradis:02}, and animal foraging movements \citep{austin:04}. In the latter case, the fit of a heavy-tailed distribution has been used as evidence that the optimal foraging strategy is a L\'evy walk with exponent $\mu\approx 2$ \citep{viswanathan:96,viswanathan:99}. However, the appropriateness of heavy-tailed distributions for some of these data sets, and the methods used to fit them, have recently been questioned by Edwards {\em et al} \citep{edwards:07}.

In a literature search of studies on foraging movements, ecological size spectra and dispersal functions, the first 24 papers that fitted power-law distributions to data were chosen \citep{passow:94,cole:95,viswanathan:96,vidondo:97,keitt:98,gin:99,viswanathan:99,camacho:01,cavenderbares:01,marell:02,paradis:02,bartumeus:03,chapman:03,quinones:03,austin:04,ramosfernandez:04,bertrand:05,katul:05,sundberg:05,weimerskirch:05,dauer:06,klein:06,brown:07,vanhoutan:07}. In only six of these 24 papers was it clear that the authors correctly fitted a statistical distribution to the data, whereas 15 used a flawed fitting procedure (and in the remaining three it was unclear). Ten of the studies compared the goodness-of-fit of the power-law distribution to other types of distribution, but at least two of these ten used comparison methods that were invalid in this context. Correctly fitting a power-law distribution to data is not difficult but requires a little more statistical knowledge than the standard linear regression and linear correlation techniques used in most cases.

In this report, the most common methods currently used to fit a power law to data are reviewed, and some of the drawbacks of these methods are discussed. A method that avoids these drawbacks, based on maximum likelihood estimation, is then described in a step-by-step guide (the reader is also referred to \citep{edwards:07}). Two case studies are presented illustrating the advantages of this method.

\section{Methods}
\subsection*{Linear regression methods}
The probability density function (PDF) for a power-law distribution has the form
\[
p(x)=c x^{-\mu} \qquad \mbox{where } \mu\ge 1.
\]
This function is not defined on the range 0 to $\infty$, but on the range $\xmin$ to infinity, otherwise the distribution cannot be normalised to sum to $1$. This condition implies that the exponent $\mu$ is related to $\xmin$ and $c$ by
\[
c=\left(\mu-1\right)x_\mathrm{min}^{\mu-1}.
\]

A common problem is to determine whether a particular sample of data is drawn from a power-law distribution and, if so, to estimate the value of the exponent $\mu$. A standard approach to this problem (used by 14 of the papers in the literature sample) is to bin the data and plot a frequency histogram on a log--log scale. If the sample is indeed drawn from a power-law distribution and the correct binning strategy is used, there is a linear relationship between $\ln x$ and $\ln p(x)$, so the frequency plot will produce a straight line of slope $-\mu$ \citep{sims:07}. Alternatively, one may use the cumulative distribution function (CDF) $C(x)$, which shows a linear relationship between $\ln x$ and $\ln\left(1-C(x)\right)$. So plotting $1-C(x)$ (\ie the relative frequency of observations with a value greater than $x$) against $x$ (called a rank--frequency (RF) plot) on a log--log scale will give a straight line of slope $1-\mu$. 

All data sets are subject to statistical noise, particularly in the tail of the distribution. Hence a frequency chart will never produce a perfectly straight line, so some kind of fitting procedure is necessary. The most common method, used by 16 of the 24 studies surveyed, is to estimate the power-law exponent by using linear regression to find the line of best fit on the frequency histogram or RF plot.

\subsection*{Drawbacks of the linear regression method}
Linear regression assumes that one is free to vary both the slope and intercept of the line in order to obtain the best possible fit. However, this is not true in the case of fitting a probability distribution, because of the constraint that the distribution must sum to $1$. Once the range of the data has determined $\xmin$, the power-law distribution has only one degree of freedom $\mu$, as opposed to the two afforded by the naive linear regression or line of best fit approach. Unless this constraint is explicitly acknowledged and respected, the fitting procedure will produce an incorrect estimate of the exponent $\mu$ and the fitted distribution will not be a PDF on the appropriate $x$ range (see for example Case Study 1).

Furthermore, the linear regression method does not offer a natural way to estimate the size of the error in the fitted value of $\mu$ (very few of the studies surveyed made any attempt to do this). A point value for $\mu$ on its own is of questionable merit. 

A third criticism of the linear regression method of fitting a power law to some sample is that very rarely is any meaningful attempt made to judge the goodness-of-fit of the proposed model. The most common approach, taken by twelve of the surveyed works, is to provide the correlation coefficient $r^2$. This is a measure of the strength of linear correlation between $\ln x$ and $\ln p(x)$, but is not a measure of the goodness-of-fit of a proposed model such as a power law. Of course, in the absence of any \textit{a priori} knowledge of the distributions of errors in the data, measuring goodness-of-fit of a single model is extremely difficult. However, it is possible to compare goodness-of-fit of two or more candidate models. In the work surveyed, nine papers explicitly compared the goodness-of-fit of the power-law distribution with other candidate models, but at least two of these nine used a comparison based on the flawed linear regression method and on $r^2$, so in reality added little to the analysis. 

Finally, although it is not necessary to bin the data to use a linear regression method, 14 of the 24 surveyed papers did so in their analyses, and nine used bins of equal width. It should be noted that, when a sample that does follow a power-law distribution is binned using fixed width bins, the log--log plot of the PDF is not linear. To see a linear relationship one must use logarithmic width bins. This fitting approach is described in detail in \citep{sims:07}, but the statistical results are often sensitive to the binning strategy (in particular the logarithmic base) used \citep{vidondo:97} and the occurrence of empty bins is problematic. Case Study 1 illustrates some of the nonsensical conclusions that can result from binning the data. By far the best approach to fitting a power law is not to bin the data in any way, but to use the raw data whenever possible.

\subsection*{The maximum likelihood method}
A simple technique for fitting a power-law distribution is to use maximum likelihood estimation (MLE). This method provides an unbiased estimate of the exponent $\mu$ \citep{newman:06}. Equally importantly, it provides an estimate of the error in the fitted value of $\mu$, and allows the goodness-of-fit of different candidate models to be directly compared \citep{burnham:02}. Furthermore, the method does not require the data to be binned in any way. In the step-by-step guide below, the method is used to compare the fit of a power-law distribution and an exponential distribution, $p(x)=\lambda e^{-\lambda\left(x-\xmin\right)}$, on the same range $(\xmin,\infty)$, but can be easily adapted to compare with other candidate distributions (\eg normal distribution, gamma distribution).

The method proceeds in the following steps.
\begin{enumerate}
\item \textbf{Draw a RF plot.} To gain a picture of the distribution of the data, plot the fraction of observations greater than $x$ against $x$ on a log--log plot by:
\begin{itemize}
\item ordering the data from largest to smallest so that $x_1\ge x_2\ge\ldots \ge x_N$;
\item plotting $\ln\left(\frac{j}{N}\right)$ against $\ln x_j$ (for $j=1,\ldots,N$).
\end{itemize}
\item \textbf{Choose $\xmin$.} In some cases, it may be desirable to examine only the `tail' of the distribution by first discarding from the sample all values less than some cut-off $x_\mathrm{min}$. Typically, this is chosen so as to disregard the curved part of the RF plot on its left-hand side. If it is required to fit a distribution to the entire sample, take $\xmin$ to be the smallest observation.
\item \textbf{Calculate MLE parameters and likelihoods:}
\begin{itemize}
\item MLE power-law exponent
\[
\mu_\mathrm{MLE}= 1+M\left(\sum_{i=1}^M \ln\frac{x_i}{x_\mathrm{min}}\right)^{-1},\]
\item log-likelihood for power-law model
 \begin{eqnarray*} \mathcal{L}_\mathrm{pow} &=& M\left(\ln(\mu_\mathrm{MLE}-1) -\ln x_\mathrm{min}\right)\\
&&- \mu_\mathrm{MLE}\sum_{i=1}^M\ln\frac{x_i}{x_\mathrm{min}},\end{eqnarray*}
\item MLE exponential parameter 
\[ \lambda_\mathrm{MLE} = M\left(\sum_{i=1}^M (x_i-x_\mathrm{min}) \right)^{-1},\]
\item log-likelihood for exponential model
 \[ \mathcal{L}_\mathrm{exp}  = M\ln\lambda_\mathrm{MLE} - \lambda_\mathrm{MLE} \sum_{i=1}^M (x_i-x_\mathrm{min}),\]
 \end{itemize}
 where $M$ is the number of data points in the (truncated) sample. See \citep{newman:06} for a derivation of these formulae.

\item \textbf{Select the best model.} For each model, calculate the Akaike information criterion (AIC) and Akaike weight ($w$), which are defined, for model $i$, by \citep{burnham:02}:
\begin{eqnarray*}
\mathrm{AIC}_i&=& -2 \mathcal{L}_i + 2K_i\\
w_i&=& \frac{\exp\left( -\frac{\mathrm{AIC}_i-\mathrm{AIC}_\mathrm{min}}{2}\right)}{\sum_{j=1}^p \exp\left( -\frac{\mathrm{AIC}_j-\mathrm{AIC}_\mathrm{min}}{2}\right)}
\end{eqnarray*}
where $K_i$ is the number of parameters in model $i$ ($K=1$ for the power-law and exponential models), $p$ is the total number of models being compared (in this case, comparing a power-law model and an exponential model means that $p=2$) and $\mathrm{AIC}_\mathrm{min}$ is the smallest of the AIC across all $p$ models.

The Akaike weight gives a measure of the likelihood that a particular model provides the best representation of the data. If $w_\mathrm{pow}>w_\mathrm{exp}$ then a power-law distribution provides a better fit to the data than an exponential distribution, and the estimated value of the exponent of the power law is $\mu_\mathrm{MLE}$. If $w_\mathrm{pow}<w_\mathrm{exp}$ then an exponential distribution is favoured over a power law on that range.

\item \textbf{Calculate the error in the estimate of $\mu$.} The standard deviation $\sigma$ of the estimated power-law exponent $\mu_\mathrm{MLE}$ may also be calculated:
\[
\sigma = \sqrt{M}\left(\sum_{i=1}^M \ln\frac{x_i}{x_\mathrm{min}}\right)^{-1}.
\]
This gives the average size of the error in the estimated value of $\mu$.

\end{enumerate}

Six of the 24 surveyed studies correctly fitted the data and compared different candidate models using a method similar to the one described above. (It is interesting to note that these six were all found in the dispersal literature.)

\section{Results}
\subsection*{Case study 1}
\begin{figure}
\begin{center}
\includegraphics[width=8cm]{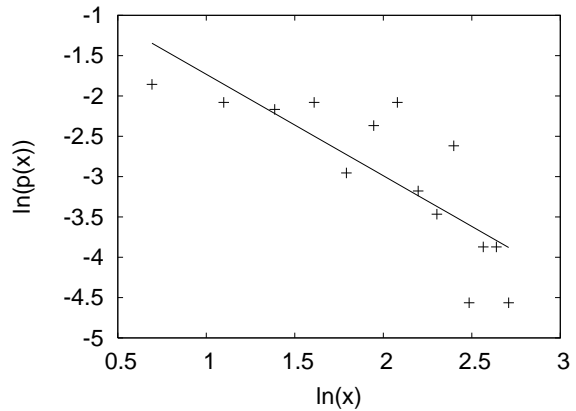}
\end{center}
\caption{Relative frequency histogram of seal movement lengths $x$ (in km) on a log--log scale, together with the line of best fit $\ln(p(x))=-1.2573\ln(x)-0.4741$. This line is not a probability distribution on an appropriate $x$ range.}
\label{fig:seals}
\end{figure}

Austin {\em et al} \citep{austin:04} fitted power-law distributions to the movement lengths of grey seals; Figure \ref{fig:seals} shows an example data set consisting of 96 observations. The original data are already binned, so it was necessary, for the re-analysis, to assume that all observations in the sample were at the midpoint of their respective bin. Using linear regression, the line of best fit has equation $Y=-1.26X -0.474$ (in log coordinates), leading to an estimated value of the exponent of $\mu=1.26$. However, it was omitted from \citep{austin:04} that, in order for the distribution to sum to $1$, this line of best fit implies an $\xmin$ value of $31$ km. As the data are in the range $2$ km to $15$ km, it is nonsensical to fit a distribution with range $31\le x\le \infty$.

These data were re-analysed in \citep{sims:07}. This work used logarithmic binning but, again, used linear regression to find the line of best fit and estimated the exponent to be $\mu=0.8$. This exponent cannot be correct as any power law with an exponent of less than $1$ cannot be normalised to sum to $1$ and hence is not a probability distribution. This illustrates the major drawbacks of using linear regression on binned data: the fitted line may be sensitive to the size and number of bins used and to the occurrence of empty bins, and may result in a power law that is not an appropriate probability distribution.

Using the maximum likelihood method outlined above, the estimated exponent is $\mu_\mathrm{MLE}=2.05$ (with an expected error of $\sigma=0.108$). This explicitly presumes a minimum data value of $\xmin=2$ km (the smallest observation in the sample). However, comparing to an exponential distribution as described above shows that the exponential distribution (with $\lambda=0.246$) provides a much better model for this data set than a power law ($w_\mathrm{pow}<10^{-7}$). Hence, the power-law hypothesis for this data set can be confidently rejected.

\subsection*{Case study 2}
\begin{figure}
\begin{center}
\includegraphics[width=8cm]{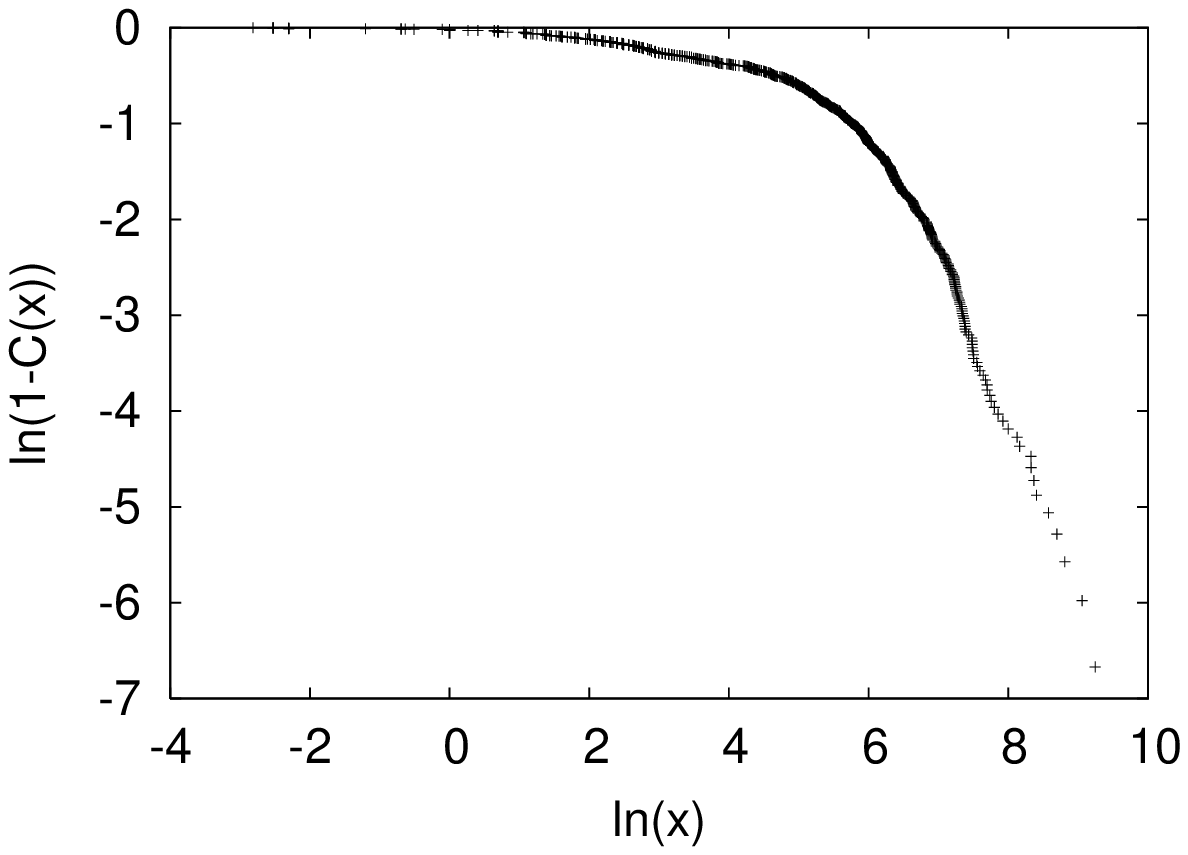}\\
 (a)\\
\includegraphics[width=8cm]{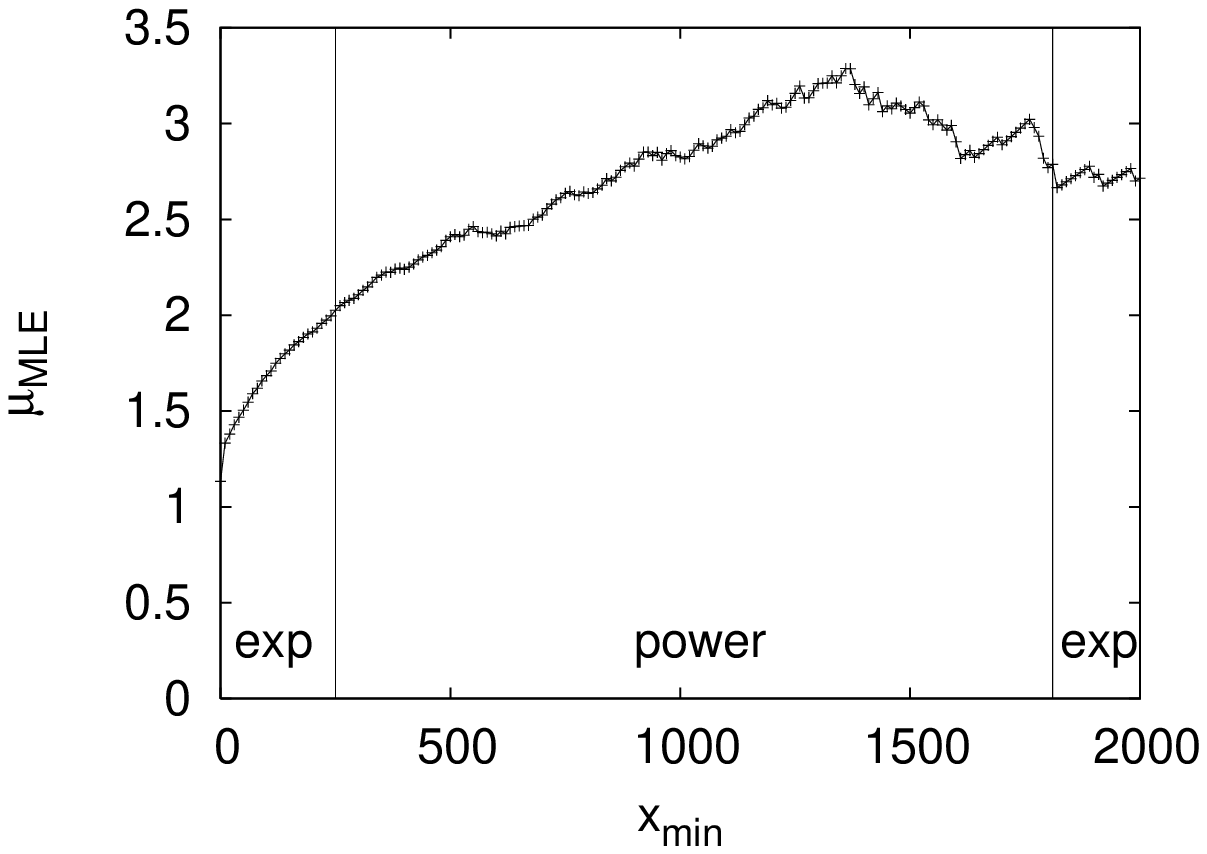}\\
(b)
\end{center}
\caption{(a) RF plot for the farm size data set. (b) Graph of the estimated exponent $\mu_\mathrm{MLE}$ against the cut-off $\xmin$. A power law provides a better fit to the data than an exponential distribution for value of $\xmin$ between $250$ and $1810$; an exponential provides a better fit for values of $\xmin$ outside this range.}
\label{fig:farms}
\end{figure}

The data set shown in Appendix \ref{sec:appb} contains the land area (in hectares) of 789 farms in the Hurunui district of New Zealand; this data set was chosen for illustrative purposes. The RF plot for this data set is shown in Figure \ref{fig:farms}(a). We find the maximum likelihood estimate of the power-law exponent for a range of values of the minimum cut-off $\xmin$ (discarding all data less than $\xmin$) using the method described above. By calculating the Akaike weights, it is possible to determine whether an exponential or a power-law distribution provides the better fit to the data, for any given value of $\xmin$. A power law is favoured over an exponential ($w_\mathrm{pow}>w_\mathrm{exp}$) for all choices of $x_\mathrm{min}$ between $250$ and $1810$, but an exponential is favoured ($w_\mathrm{pow}<w_\mathrm{exp}$) for $\xmin<250$. Figure \ref{fig:farms}(b) shows the estimated power-law exponent $\mu_\mathrm{MLE}$ for the different $\xmin$ values: it is clear that the estimated exponent has a wide range of values from $2$ to $3.3$, depending on the range of data used. It should also be noted that, if the aim is to find the best statistical model describing the entire data set (\ie $\xmin$ equal to the smallest observation in the sample), then an exponential distribution (with $\lambda=2.33\times 10^{-3}$) is clearly favoured over a power law ($w_\mathrm{pow}<10^{-65}$).
The dependence of the outcome of the analysis on the arbitrary selection of $\xmin$ further illustrates the potential problems with fitting power-law distributions.

\section{Discussion}
Some of the statistical procedures commonly used to fit power-law distributions to ecological data often lead to misleading or incomplete conclusions \citep{edwards:07}. In this paper, a simple step-by-step method for fitting a power law to a data set has been described. This method, based on maximum likelihood estimation, provides an unbiased estimate of the power-law exponent, as well as the expected error in this value, and assesses the goodness-of-fit of a power-law model compared to alternative candidate models such as the exponential distribution.

\section*{Acknowledgements}
The authors are grateful to Environment Canterbury for providing the data set on farm sizes, and to Richard Duncan and two anonymous referees for useful comments that helped to improve this paper.

\appendix
\section{Farm size data}
\label{sec:appb}
The following data set (courtesy of Environment Canterbury) contains the sizes (in hectares) of 789 farms in the Hurunui district of New Zealand.

\footnotesize
0.06, 0.08, 0.08, 0.08, 0.1, 0.1, 0.1, 0.1, 0.1, 0.3, 0.5, 0.5, 0.53, 0.6, 0.9, 1, 1, 1, 1, 1, 1, 1, 1.3, 1.5, 1.9, 1.9, 2, 2, 2, 2, 2, 2, 2, 2, 2, 2, 2.3, 2.9, 2.9, 2.9, 3, 3, 3, 3, 3, 3, 3, 3, 3.2, 3.3, 3.5, 3.8, 3.9, 4, 4, 4, 4, 4, 4, 4, 4, 4, 4.2, 4.2, 4.3, 4.6, 4.8, 4.9, 5, 5, 5, 5, 5, 5.4, 5.7, 6, 6, 6, 6, 6, 6, 6.2, 6.3, 6.3, 7, 7, 7, 7, 7, 7.2, 7.5, 7.9, 8, 8, 8, 8, 8, 8, 8, 8.3, 8.8, 9, 9, 9, 9, 9.3, 9.9, 10, 10, 10, 10, 10, 10.2, 10.4, 11, 11, 11, 11, 11, 11.8, 12, 12, 12, 12, 12, 12, 12, 12, 12.1, 13, 13, 1, 13, 13.8, 13.9, 14, 14, 14, 14, 14.4, 14.5, 15, 15, 15, 15, 15, 15, 15.2, 15.7, 15.8, 16, 16, 16, 16, 16, 17, 17, 17, 17, 17, 17.1, 17.1, 17.5, 17.6, 18, 18, 18, 18, 18, 18, 18, 18, 19, 19, 19, 19, 19, 19, 19, 19, 19, 20, 20, 20, 21, 21, 21, 21.9, 22, 22, 23, 23, 23, 24, 24, 24, 25, 25, 26, 27, 27, 27, 28, 28, 29, 29, 30, 31, 31, 32, 32, 32, 32, 32, 32, 32.9, 34, 34, 35, 35, 35, 36, 36, 36, 37, 38, 38, 39, 39, 39, 40, 41, 43, 43, 44.1, 45, 45, 46, 46, 46, 46, 47, 47, 47, 47, 48, 49, 50, 53, 54, 55, 56, 56.5, 57, 58, 58, 60, 63, 67, 68, 69, 70, 71, 72, 72, 72, 73, 73, 74, 75, 75, 75, 75, 77, 78, 78, 78, 78, 79, 80, 81, 82, 83, 85, 85, 87, 87, 90, 90, 91, 91, 91, 92, 93, 93, 96, 9, 97, 99, 99, 99, 100, 100, 100, 101, 101, 102, 103, 103, 104, 104, 105, 106, 107, 109, 109, 113, 115, 117, 118, 120, 121, 121, 121, 124, 124, 124, 124, 125, 126, 12, 128, 128, 131, 131, 132, 133, 135, 135, 136, 136, 137, 137, 137, 138, 141, 142, 143, 143, 143, 143, 144, 144, 145, 145, 148, 148, 149, 149, 150, 151, 153, 155, 15, 156, 157, 157, 158, 160, 161, 162, 162, 162, 163, 164, 164, 165, 167, 167, 169, 169, 170, 171, 171, 172, 173, 173, 173, 175, 176, 177, 181, 182, 182, 183, 183, 18, 184, 185, 186, 188, 188, 190, 191, 192, 192, 192, 193, 193.5, 194, 195, 196, 197, 198, 198, 200, 200, 200, 202, 202, 203, 206, 207, 209, 210, 210, 213, 216, 216, 219, 221, 223, 223, 226, 226, 226, 228, 229, 229, 230, 232, 233, 233, 234, 237, 239, 242, 245, 247, 247, 250, 250, 252, 254, 256, 258, 263, 267, 267, 267, 268, 269, 269, 269, 270, 271, 273, 273, 273, 275, 277, 277, 279, 280, 282, 283, 284, 284, 284, 288, 289, 289, 295, 295, 298, 298, 298, 299, 301, 305, 306, 308, 309, 311, 31, 313, 313, 316, 316, 320, 322, 324, 327, 332, 333, 333, 339, 341, 342, 345, 346, 348, 349, 349, 352, 353, 354, 354, 355, 357, 361, 362, 364, 365, 365, 365, 366, 36, 367, 372, 372, 375, 376, 379, 380, 384, 384, 385, 388, 388, 388, 390, 391, 392, 394, 395, 395, 396, 398, 399, 401, 401, 401, 404, 405, 410, 410, 415, 416, 423, 42, 428, 430, 431, 431, 432, 440, 443, 443, 448, 448, 450, 453, 456, 457, 461, 461, 464, 466, 471, 476, 478, 480, 496, 497, 498, 502, 504, 504, 505, 511, 512, 513, 51, 515, 517, 518, 520, 524, 526, 529, 533, 540, 542, 546, 550, 550, 552, 552, 553, 555, 555, 558, 563, 565, 565, 567, 567, 571, 571, 576, 578, 580, 581, 583, 587, 58, 591, 597, 597, 597, 598, 602, 611, 612, 614, 614, 616, 630, 635, 635, 640, 640, 645, 650, 652, 652, 661, 662, 667, 681, 685, 695, 699, 719, 720, 731, 732, 741, 75, 752, 760, 763, 764, 769, 771, 774, 776, 785, 790, 792, 792, 804, 807, 814, 824, 843, 844, 849, 859, 874, 886, 892, 892, 892, 923, 923, 932, 935, 938, 942, 952, 95, 955, 958, 975, 980, 984, 989, 996, 999, 1002, 1004, 1013, 1040, 1044, 1054, 1059, 1064, 1082, 1094, 1110, 1117, 1124, 1153, 1172, 1191, 1196, 1201, 1216, 1219, 1226, 1261, 1262, 126, 1278, 1303, 1312, 1336, 1339, 1364, 1371, 1372, 1379, 1380, 1387, 1401, 1403, 1405, 1431, 1431, 1439, 1459, 1477, 1489, 1491, 1521, 1534, 1537, 1544, 1560, 176, 1593, 1594, 1607, 1608, 1631, 1690, 1765, 1778, 1780, 1780, 1798, 1812, 1815, 1899, 1917, 1986, 2081, 2152, 2200, 2205, 2288, 2335, 2440, 2576, 2758, 2975, 3378, 3512, 4126, 4136, 4300, 4466, 5299, 5967, 6700, 8596, 10336.

\end{document}